# Ultra-Low Noise Balanced Receiver with >20 dB Quantum-to-Classical Noise Clearance at 1 GHz

D. Milovančev, F. Honz, N. Vokić, F. Laudenbach, H. Hübel, B. Schrenk

AIT Austrian Institute of Technology, GG4, 1210 Vienna, Austria. dinka.milovancev@ait.ac.at

**Abstract** *We demonstrate a die-level balanced homodyne receiver for coherent optical access and continuous-variable quantum applications, featuring a 40dB CMRR up to 1GHz and a high quantum-to-classical noise ratio of 26.8dB at 12.3mW of LO power. 500Mb/s QPSK transmission was accomplished with a sensitivity of -55.8dBm.*

## Introduction

The need for coherent receivers is fuelled by the recent industrial discussion on the introduction of coherent detection in short reach networks and, on the longer term, in the optical access segment. On top of this, quantum technologies promise performance gains in the fields of sensing and computation, while equipping our communication infrastructure with information-theoretic secure crypto primitives. Continuous-variable (CV) schemes[1,2] build on coherent states to encode information and highly sensitive balanced homodyne detectors (BHD). A technology overlap therefore exists with coherent optical telecommunications, especially with respect to low-bandwidth receivers with unprecedented noise performance. Earlier works on coherent optical access have demonstrated impressive sensitivities of -53 dBm for 311 Mb/s data rates[3]. However, the receiver performance that is sought for a multi-purpose use of GHz-class coherent receivers for access and quantum applications must address noise before bandwidth. This is underpinned by the large electronic noise contribution in highly sensitive balanced receivers: it typically amounts to ~65% of excess noise[4] and renders the opto-electronic integration as the main challenge for further performance improvements.

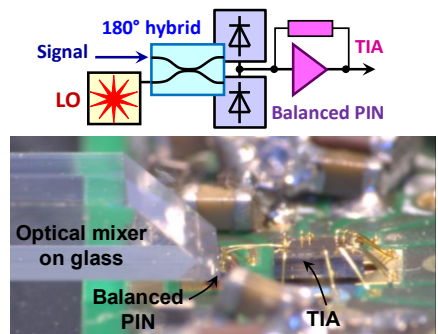

**Fig. 1:** Die-level subassembly of the CV receiver.

In this work, we are pursuing a die-level balanced receiver that achieves both, a large bandwidth of 750 MHz and a high quantum-to-classical noise ratio (QCNR) of 26.82 dB at 12.3 mW of local oscillator (LO) power. We discuss the performance offerings for CV-QKD applications and experimentally demonstrate the multi-purpose use of the receiver for 500 MB/s QPSK transmission with a sensitivity of -55.8 dBm.

## Multi-Purpose Balanced Receiver

One of the prime requirements for a BHD are a high QCNR to minimize the excess noise and to operate in the shot-noise limited regime. This noise performance is paired with a high common mode rejection ratio (CMRR) of typically 30 dB,[5] in order to suppress LO noise and adjacent channel beat terms. There is increasing need for an extended bandwidth, with coherent access aiming at unshared 1 Gb/s/λ per-user rates and quantum communications striving for >100 MHz reception bandwidth.

The direct adoption of available telecom-grade BHDs is not suitable since these are speed optimized (>10 Gb/s) rather than noise optimized. Recent works therefore focus on custom BHD receivers, with most of these building on discrete fibre-optics and opto-electronic components. Due to the associated parasitics of the assemblies, relatively modest bandwidths of up to 250 MHz[6] and 300 MHz[7] can be accomplished under acceptable noise performance targets. The highest reported bandwidth using discrete photodiodes that are ac-coupled to a 50Ω RF-IC front-end was 1.2 GHz,[8] featuring a QCNR of 18.5 dB at a LO power of 8.8 mW. The 50Ω-matching of photodiodes and RF-IC suits the bandwidth but does not yield best noise performance. Besides the noise-bandwidth trade-off, the use of ROSA-packaged photodiodes off-loads the optical mixing of LO and signal to fibre-optic components, which makes the power balancing and skew minimization more difficult.

In order to unlock performance gains, a shift toward die-level receiver assemblies is needed. Recently, squeezed light detection based on a PIC-integrated BHD with wire-bonded transimpedance amplifier (TIA) was reported,[9] achieving a bandwidth of 1.7 GHz with a clearance of 14 dB at 4.36 mW. The optical mixer on PIC further ensures a good matching of the photodiodes. Here, we are pursuing a die-level receiver with an optical mixer inscribed on glass. The mixer is vertically coupled to a PIN array with low-noise TIA circuit so to ensure good photodiode matching, skew-free operation and good optical power balancing.

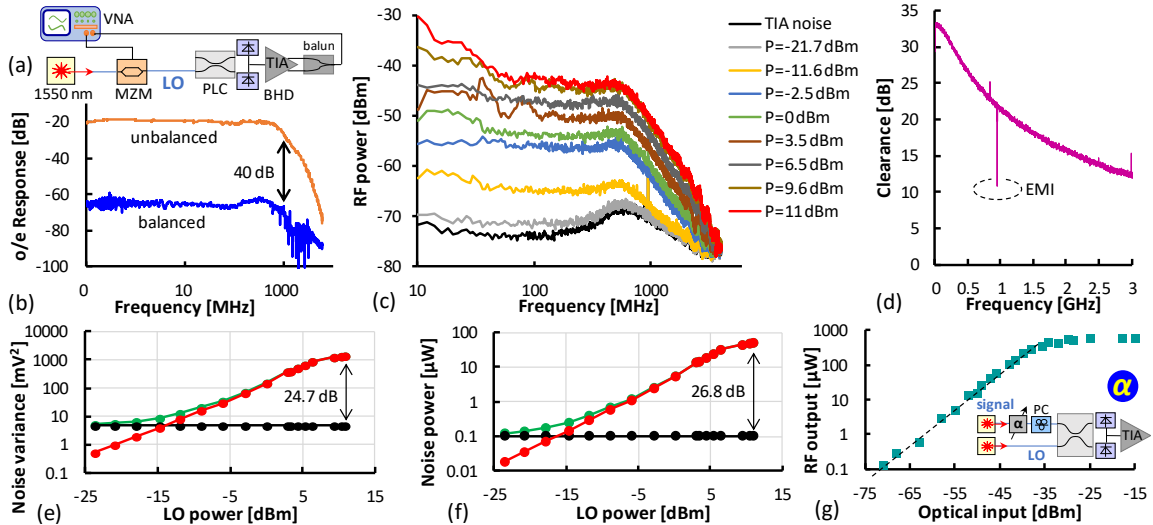

**Fig. 2:** (a) CMRR characterization setup. (b) CMRR response. (c) Received power for dark and lit LO. (d) Clearance for an LO power of 10.9 dBm and noise performance: (e) time-domain QCNR and (f) frequency-domain QCNR. (g) TIA linearity.

**Die-Level Balanced Receiver Assembly**
In order to minimize the large electronic excess noise contribution of the balanced receiver, a low-noise TIA rated for 1.25 Gbit/s data rates and a very low input-referred rms noise current of 60 nA was hybrid co-integrated with a balanced PIN detector (Fig. 1). Since the TIA noise scales with the square root of photodiode capacitance $\sqrt{C_{PD}}$,[10] we chose the commercial die level PIN array with $C_{PD}$ of 200 fF. This is much lower than that of packaged photodiodes, which can reach 1 pF. The two photodiodes in the form of a linear array were wire-bonded to the TIA and to their positive and negative biasing branches. Since the photodiodes are on the same chip and have P/N bonding pads, a high level of matching can be expected. The photodiode responsivity is 1.0 A/W at 1550 nm. The photodiode pitch of 250 µm and the diameter of 40 µm ensure compatibility with the vertical coupling to the glass-inscribed optical mixer. This planar lightwave circuit (PLC) constitutes the optical 180° hybrid required for simplified coherent receivers in optical access, or homodyne CV receivers, as they are commonly adopted in QRNG and CV-QKD systems.[2] The vertically coupled glass interposer ensures that cladding modes of the injected LO are not illuminating the balanced detector. Signal and LO inputs are pigtailed to the PLC. The coupling efficiency between PLC and photodiode was 80 to 85%.

**Characterization of the Receiver**
*CMRR:* the frequency-dependent CMRR has been determined through S-parameter measurement, using an amplitude modulated optical input at 1550 nm (Fig. 2a). The optical input power was low enough to make sure that there is no saturation in the case of unbalanced detection – single photodiode illumination, for which all the dc photocurrent is sinked or sourced by the TIA. Balancing has been accomplished through lateral alignment of the PLC with respect to the photodiode array. Under balancing, the photocurrent for each photodiode was 8.4 µA for an input power of -17 dBm. At 1 GHz the CMRR is as high as 40 dB (Fig. 2b).

*QCNR:* The quantum-to-classical clearance was evaluated under detector balancing for an unmodulated light injection through the LO. The corresponding power spectra (Fig. 2c) for various optical LO levels show the rise of clearance over a broad frequency range up to the TIA bandwidth. The corresponding clearance reaches a value of 21.5 dB at 1 GHz for a LO power of 10.9 dBm (Fig. 2d). In the time domain, the QCNR can be estimated based on time trace noise variance (Fig. 2e). We estimated the electronic (TIA only) noise variance (black) and total noise variance (green). The quantum noise variance is obtained by subtracting the electronic noise variance from the total noise variance (red). At an LO power of 10.9 dBm, we obtained a QCNR of 24.74 dB. At LO levels beyond 9.4 dBm the receiver enters the saturation region and there is little improvement in QCNR. By integrating the frequency-dependent noise from 1 MHz up to 1 GHz at each LO power, we can get an estimate for the QCNR in the frequency domain. Figure 2f shows that at the maximum LO power the

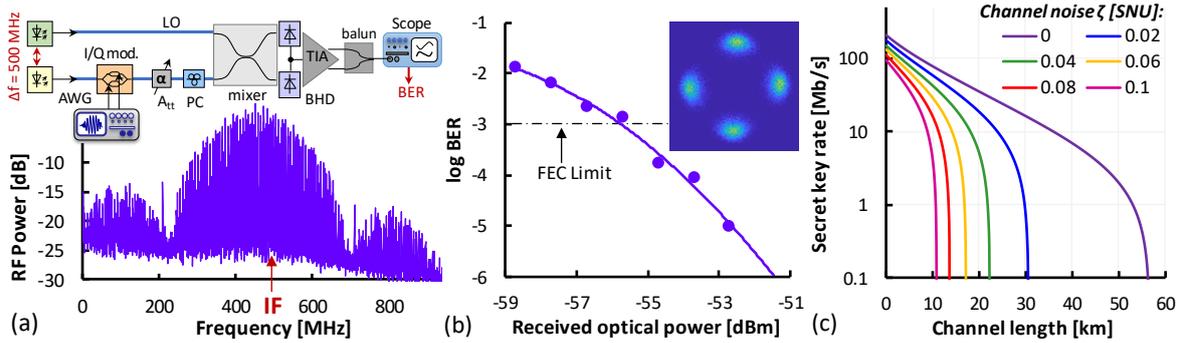

**Fig. 3:** (a) Heterodyned QPSK after coherent detection. (b) BER performance for 500 Mb/s QPSK reception. (c) Simulated secure-key rate for CV-QKD with untrusted receiver, based on the noise characteristics of the demonstrated CV receiver.

QCNR is 26.8 dB.

*TIA linearity:* We further investigated the linear operation region of the receiver through power analysis for single-tone reception. This tone has been generated by beating the LO with an optical input signal (Fig. 2g, α) whose wavelength was shifted by 120 MHz. The LO power was kept fixed at 100 µW while the signal power has been varied. Figure 2g shows the linear range, which spans from -71 dBm up to -38 dBm, therefore yielding a dynamic range of 33 dB. Since the mixing product of the signal that is being post-amplified by the TIA is proportional to $\sqrt{(P_{LO}P_{sig})}$, the maximum permissible signal power can be estimated for different LO power levels.

The performance offerings of the receiver have been evaluated for coherent access and CV-QKD applications. Coherent heterodyne reception of a complex (i.e. I/Q modulated) optical signal at an intermediate frequency (IF) of 500 MHz has been considered, with the required 90° hybrid needed for I/Q demodulation being implemented in the digital domain.

## Coherent Receiver for Optical Access

The bandwidth of the receiver allows for the simplified heterodyne reception of complex signals with unshared per-user data rates in the 1-Gb/s range. Coherent heterodyning at an IF, together with the optical mixer inscribed on the PLC, can mitigate the loss and the complexity of an optical 90° hybrid by off-loading I/Q demodulation to the DSP domain. We experimentally verified the reception of a 250 Mbaud QPSK signal with our single-polarization receiver at an IF of 500 MHz (Fig. 3a). The BER performance for an LO power of 13 dBm is presented in Fig. 3b. The sensitivity at the hard-decision FEC threshold of $1 \cdot 10^{-3}$ is -55.8 dBm. This sensitivity allows for a high optical budget of 49.8 dB, considering a low multi-channel launch of -6 dBm/λ from the central office.

## CV-QKD with Untrusted Receiver

The performance of the low-noise receiver for CV-QKD has been evaluated based on simulations[4]. Homodyne quantum measurement is considered for a 250 Mbaud, Gaussian modulated signal for which the modulation variance was dynamically optimized. We chose a channel loss of 0.23 dB/km and a total detection loss of 1.2 dB due to net responsivity (including vertical coupling) and the excess loss of the optical 180° hybrid used for photomixing of signal and LO. The reconciliation efficiency was 0.97 and an asymptotic key rate was considered. The noise performance of the receiver of this work leads to a receiver excess noise of 0.00336 shot-noise units (SNU). Figure 3c reports the secure-key rate (SKR) that can be expected for various values of the channel noise ζ, the latter being referred to the receiver input.

A high SKR of 43 Mb/s can be expected for a reach of 10 km and ζ = 0.04 SNU. Short-reach networks can leverage on a high SKR of more than 100 Mb/s. For a reduced ζ of 0.02 SNU, a reach of up to 29.8 km is supported at a SKR of 1 Mb/s. It shall be noted that the provided results in Fig. 3c are considering an untrusted receiver scenario. This means that the entire receiver noise is considered in the security analysis, since an eavesdropper can potentially manipulate this noise component.

## Conclusions

We have demonstrated a low-noise BHD as a die-level subassembly for coherent applications in optical access and quantum communications. Its high QCNR of 26.8 dB over a bandwidth of 1 GHz allows a high SKR of 100 Mb/s for short-reach applications, while further allowing for classical 500 Mb/s QPSK reception with a sensitivity of -55.8 dBm. LO co-integration with its optical mixer on glass is left for future work.

## Acknowledgements

This work was supported in part by the ERC under the EU Horizon-2020 programme (grant n° 804769) and by the Austrian FFG Agency (grant n° 884443).